# Topological defects at smectic interfaces as a potential tool for the biosensing of living microorganisms

Vajra S. Badha,[a] Tagbo H.R. Niepa,[b,c] and Mohamed Amine Gharbi[a]*

[a] Department of Physics, University of Massachusetts Boston, Boston, MA 02125, USA

[b] Department of Chemical Engineering, Carnegie Mellon University, Pittsburgh, PA 15213, USA

[c] Department of Biomedical Engineering, Carnegie Mellon University, Pittsburgh, PA 15213, USA

*Corresponding author: Mohamed.Gharbi@umb.edu

**ABSTRACT**

Characterizing the anchoring properties of smectic liquid crystals (LCs) in contact with bacterial solutions is crucial for developing biosensing platforms. In this study, we investigate the anchoring properties of a smectic LC when exposed to *Bacillus Subtilis* and *Escherichia coli* bacterial solutions using interfaces with known anchoring properties. By monitoring the optical response of the smectic film, we successfully distinguish different types of bacteria, leveraging the distinct changes in the LC's response. Through a comprehensive analysis of the interactions between bacterial proteins and the smectic interface, we elucidate the potential underlying mechanisms responsible for these optical changes. Additionally, we introduce the utilization of topological defects; the focal conic domains (FCDs), at the smectic interface as an indicative measure of the bacterial concentration. Our findings demonstrate the significant potential of smectic LCs and their defects for biosensing applications and contribute to our understanding of bacteria-LC interactions, paving the way for advancements in pathogen detection and protein-based sensing.

# 1. INTRODUCTION

A biological sensor detects substances such as bacteria,[1,2] viruses,[3,4] and other molecules[5] impacting quality of life. Their use is ubiquitous in healthcare, where monitoring biomarker levels such as glucose[6,7] or proteins[8,9] indicative of numerous infections and diseases, is vital. In the food industry, they are used in quality control to quantify the acid levels and extent of fermentation.[10,11] In industrial and environmental applications, they are utilized to detect hazardous gases such as methane and other pollutants in coal mines and the atmosphere, respectively. Therefore, the research to develop biosensors has gained immense traction since their invention in the 1960s.[12] At its core, the working of a biosensor is simple, i.e., when a sample containing the substance of interest, the analyte is introduced to the biosensor, it interacts with the sensing agent to elicit a response to be interpreted as a result. This result can either be a simple positive/negative endpoint (qualitative) or a quantifying value (quantitative), depending on the extent of the response. For example, a BinaxNOW rapid test kit from Abbott Laboratories[13] can be used to detect the presence of COVID-19 virus based on antigen specific binding, whereas a Polymerase Chain Reaction (PCR)[14,15] test can be used to identify and quantify the presence of the COVID-19 virus based on nucleic acid sequences.

Viral and bacterial infections impose a tremendous burden[16] on the healthcare system. In many cases, delayed identification of the pathogen results in delayed or inappropriate treatment, inevitably leading to avoidable deaths. Currently, the techniques most widely used for detecting and quantifying pathogens are plate culturing,[17] PCR, Enzyme-Linked Immunosorbent Assay (ELISA),[18] and other advanced molecular assays.[9] Although these techniques are accurate, highly reliable, and detect a wide gamut of pathogenic organisms, they are time-consuming, resource-intensive, use labeled molecules, and require highly skilled technicians. Therefore, simple, quick, and reliable low-cost alternatives that can be read by non-experts are essential. In this respect, point-of-care diagnostic devices have garnered increased attention in recent years due to their advantages. Especially in the past two decades, the research to develop liquid crystal (LC) biosensors[19–22] has been rising due to their desirable biosensing abilities.

LCs are a versatile class of materials that possess properties between those of liquids and those of solids. They can flow like fluids while maintaining some molecular orientations like crystals.[23] LCs are usually made of elongated molecules that can align along particular directions. Depending on these directions and the arrangement of molecules, one can distinguish different phases of LCs. Some examples include the nematic phase, in which molecules are oriented along a common direction, known as the director; the cholesteric phase, where molecules are assembled into helical structures; and the smectic phase, in which molecules form layered structures. These different arrangements give the LCs unique optical properties making them useful in a variety of applications ranging from LC displays (LCDs)[24,25] and smart windows[26] to optical filters[27] and thermometers.[28]

LCs are also valuable for biosensing applications due to their ability to respond rapidly to changes in their environments with very high sensitivity. When an LC is in contact with a biological material, it can adjust its optical properties by changing its color or brightness due to its birefringence. This transformation can be exploited to detect the existence of analytes and measure their concentrations. An advantage of using LCs in biosensing is that they can be easily tailored to interact with specific target molecules to increase their specificity. Additionally, they are relatively simple and inexpensive to fabricate. For this reason, they were used in a wide range of applications, including in the detection of pH,[29] glucose,[30] enzymatic activity,[31] chemical[32–34] and other biochemical compounds.[35] Other applications include environmental monitoring,[36,37] food safety,[38] medical diagnostics,[39] drug discovery,[40] and detection of pathogens.[41]

The common molecular orientation of LCs can be realigned parallel or perpendicular to the surface in contact.[42] This property is called anchoring and is very well studied for various LC phases.[43,44] A planar alignment is usually obtained when the LC molecules align parallel to the surfaces confining the LC film. It could be either uniform or degenerate. However, a homeotropic alignment is achieved when the LC molecules align perpendicular to the surfaces in contact. It is also possible to create a hybrid texture when the LC anchoring at the confining surfaces is different; for example, planar on one side and perpendicular on the other one.[45] Each alignment can be distinguished by a particular optical response between crossed polarizers. By exploiting these optical properties, previous studies introduced the working principle[46–50] behind the most recent LC biosensor technology, where the analytes come into contact with the LC and elicit a response by causing an orientational reorganization of their molecules. The result is an optical response that can be visualized using imaging techniques. This principle was utilized by Popov *et al.*[51] and Pani *et al.*[52] to prove the viability of LCs as effective sensing systems.

Although many studies show the ability of LC-based systems to sense multiple analytes with specificity, as mentioned so far, most of these studies focused on nematic,[53,54] cholesteric,[55,56] and blue phase LCs,[57] as the element for sensing. Additionally, these systems only exploited the optical birefringence of the LC[58] as an output to be transduced. Little is known about the potential of other LC phases and their topological defects in biosensing applications, such as smectic LCs, despite the previous literature supporting the fact that these materials are much more sensitive to analytes than other LC phases.[44,59]

In this work, we demonstrate the ability of smectic LCs to detect the existence of living microorganisms. We illustrate how their defects, the focal conic domains (FCDs), can distinguish between various cells and also approximate their concentrations. We also discuss the interaction of living microorganisms with the smectic interface and explain how this system can be exploited in biosensing applications to detect pathogens.

## 2. MATERIALS AND METHODS

2.1. Preparation of the smectic film

Our experimental system studies the interaction between two types of living microorganisms, *Bacillus subtilis* (*B. subtilis* 6051, ATCC) and *Escherichia coli* (*E. coli* 700926, ATCC), with a smectic film confined in cross shaped polydimethylsiloxane (PDMS) holes, measuring 300 μm wide and 75 μm deep, under controlled anchoring conditions, as shown in **Figure 1-a**. The PDMS material was utilized because it is biologically inert,[60] non-toxic[61] and can be easily fabricated into microstructures. The PDMS crosslinking solution was prepared by mixing the Silicone elastomer and crosslinking curing agent (Sylgard 184 Silicone Elastomer Kit, Dow) in the ratio of 10:1. The experimental cells were first fabricated using photolithography (Microlight 3D) by printing the negative of the required microstructures with the photoresist (AZ125nXT 10A, Micro chemicals) onto a silicon substrate. After completing the photolithography process, a crosslinking PDMS mold was poured onto these microstructures and baked at 70 °C for 90 minutes.

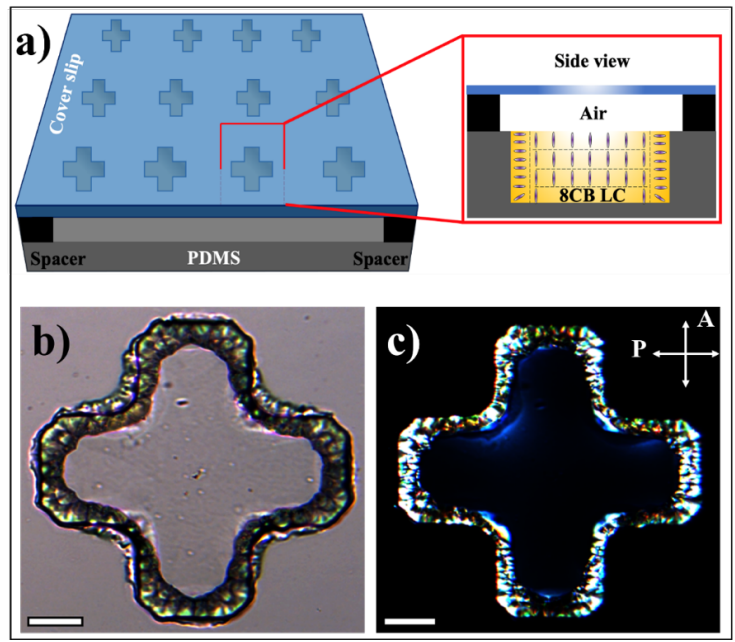

**Figure 1.** Experimental setup and anchoring property of the smectic in contact with PDMS and Air. (a) Experimental setup used to confine the smectic film into cross shaped holes made of PDMS. The side view shows the homeotropic alignment of the smectic molecules when confined between the PDMS and air. (b) Bright filed and (c) polarizing optical microscopy (POM) images of the smectic film between the PDMS and air. The scale bars are 50 μm.

The LC material utilized in this study is the 4-octyl-4′-cyanobiphenyl (8CB, Sigma Aldrich), which exhibits four phases depending on the temperature: a crystal phase below 21.35 °C, a smectic A phase between 21.35 °C and 33.35 °C, a nematic phase between 33.35 °C and 40.35 °C, and an isotropic phase above 40.35 °C.[62] We choose the 8CB because this material presents a smectic A phase at room temperature, in which the molecules are arranged in parallel layers but their director is perpendicular to the plane of these layers. Smectic LCs are characterized by their high viscosity and strong molecular order, which makes them useful for many applications including biosensing.[63,64] Although the 8CB is toxic for living microorganisms, it does not come into direct contact with bacteria due to the interfacial hydrodynamic interactions that prevent the cells from penetrating into the 8CB.[65]

The 8CB is added to the PDMS holes in its smectic phase using a spatula and subsequently by sweeping a glass cover slip over the holes to pack the LC. Further, the LC is heated to the isotropic phase using a heat gun to reduce viscosity and form a uniform layer before allowing it to cool down to room temperature. In our experiments, we tested different PDMS shapes with different sizes and choose to work with the cross shaped holes because they were found to help stabilize the smectic films in contact with the aqueous solutions instead of adding surfactants. Our goal here is to make sure that the optical response of the smectic interface is only due to its interaction with the bacterial solution and not other chemicals.

2.2. Confining living microorganisms at the smectic interface

The model microorganisms B. subtilis (6051, ATCC) and *E. coli* (700926, ATCC) were selected to represent a wide range of bacteria with different physical, physiological, and chemical properties. While the *E. coli* is Gram-negative and 1-2 μm long,[66] *B. subtilis* is gram-positive and measures 2-6 μm.[67] *E. coli* has a prominent outer membrane but *B. subtilis* cell lacks it.[68] As a result, *B. subtilis* secretes proteins and other cellular molecules directly into its surrounding environment.

Terrific Broth (TB, Sigma Aldrich) was used to grow the overnight cultures (12-14 hours) of *B. subtilis* and *E. coli* at 33°C and 37°C in the incubator (MaxQ 4450, Thermo Scientific), respectively. Dry powder of the bacteria was added into ~10 mL TB in a sterile 25 mL conical flask and placed into a shaking incubator at appropriate temperatures and 80 rpm. The growth was confirmed by a manual optical observation of turbidity of the TB solution.

Once the smectic LC layer was established, deionized water, TB with, and without microorganisms were introduced on separate samples to study their effect on the smectic anchoring. 15 μl of the culture media without washing was added to the top of LC and covered with a glass coverslip after placing 25 μm spacers. Imaging was performed from the top using a Leica DM6M microscope and recorded with a Leica DMC5400 high resolution camera. Images were collected in brightfield and polarized optical microscopy (POM) and were transferred onto FIJI for further analysis. This allowed us to visualize the effects of introducing microorganisms on the LC anchoring and investigate if smectics can serve as potential biosensors for the detection of bacteria.

2.3. Protein identification in the bacterial solutions

To better understand the interaction between the smectic interface and *E. coli* or *B. subtilis* cells, we used a Bruker tims-TOF HT mass spectrometer to identify the proteins in their solutions. Overnight cultures of bacteria were centrifuged at ~ 3250 g for 10 minutes, and 1 mL of supernatant was separated for further preparation. The proteins were precipitated out from the supernatant by adding 4 ml of acetone to the 1 ml of the supernatant and centrifuged at 2300 g to separate the proteins. Next, the pellets were resuspended in a 1 mL Optima water and buffer exchanged using pre-rinsed Amicon 3 kDa MWCO filters into Optima water before centrifuging thrice at 16000 g. The eluants were brought to 200 μL, and Bicinchoninic acid (BCA) assay was performed to determine protein concentration (see **Supplementary Information**).

## 3. RESULTS AND DISCUSSION

3.1. Anchoring properties of the smectic in contact with air, water, and TB

To study how smectic 8CB reacts to different anchoring properties, we conducted an analysis to observe its texture when in contact with PDMS and air, PDMS and deionized water, and PDMS and TB. This analysis aimed to identify the various molecular alignments that the smectic can adopt before bacteria are added. To do this, we prepared the PDMS sample with 8CB, as described in the experimental section. Next, we added 25 μm spacers on top of the PDMS and covered the setup with a glass coverslip, as shown in **Figure 1-a**. **Figures 1-b and 1-c** display the corresponding brightfield and POM images of the smectic film in this configuration. These images revealed the presence of a dark region that persists while rotating the polarizers, confirming that the 8CB molecules are perpendicular to both surfaces, the PDMS [69] and air. [70]

When deionized water is used instead of air, as shown in **Figure 2-a**, a different texture is observed both in bright field and between crossed polarizers. The dark regions that indicate homeotropic alignment disappear and are replaced by bright areas decorated with smectic defects known as FCDs (**Figures 2-b** and **2-c**). These defects consist of layers wrapping around an ellipse and a hyperbola, which contain all singular points.[71–73] The size of these defects depends on the thickness of the LC film and the curvature at the boundaries.[74] The formation of FCDs suggests that the smectic film has a hybrid anchoring — planar on one surface and perpendicular on the opposite surface. As the PDMS imposes a homeotropic alignment, therefore, the smectic has a planar anchoring in contact with water.

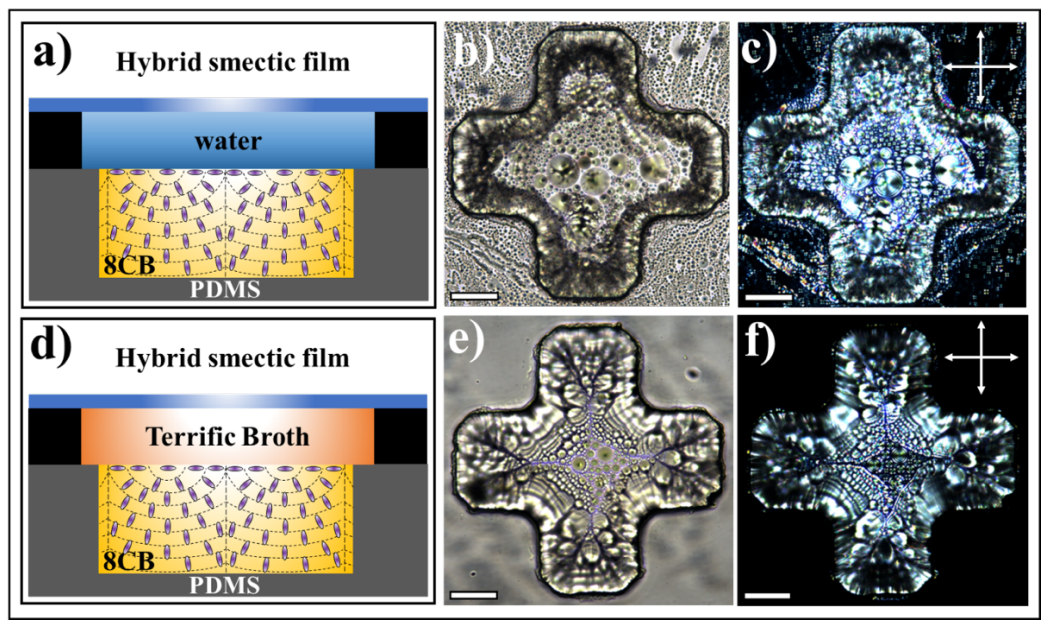

**Figure 2**. Interaction of the 8CB with water and TB. (a) Side view sketch of the LC confined between PDMS and water. Top view of the optical responses of the LC confined between PDMS and Water under brightfield (b), and POM (c). (d) Side view sketch of the LC confined between the PDMS and TB. Top view of the optical responses of the LC confined between the PDMS and TB under brightfield (e) and POM (f). The scale bars are 50 μm.

Next, we set up the smectic LC in contact with the TB instead of water, as shown in **Figure 2-d**. We chose to test the LC anchoring in contact with the TB because this fluid was used to grow *B. subtilis* and *E. coli* cells. The optical images demonstrate a flat and stable LC-TB interface decorated with FCDs, as presented in **Figures 2-e** and **2-f**. The formation of FCDs confirms that the anchoring of the 8CB is planar in contact with the TB, similar to that with water. These results also indicate that the smectic interface is sensitive to the fluids it contacts. This information can be used to explore how the smectic behaves when in contact with bacterial cultures.

3.2. Interaction of *B. subtilis* and *E. coli* cultures with a smectic interface

To understand how smectic LCs interact with living microorganisms, we first introduced a TB solution with *B. subtilis* bacteria on the top of the 8CB film, as depicted in **Figure 3-a**. Our results show that the smectic film forms a texture without any topological defects (see **Figure 3-b**), unlike pure TB, which has

FCDs at the interface. Additionally, the smectic film remains dark when observed between crossed polarizers and rotated (see **Figure 3-c**). This suggests that the 8CB has a homeotropic alignment when in contact with the TB solution containing *B. subtilis* bacteria. Subsequently, when we repeat the experiment with *E. coli* cells instead of *B. subtilis* (see **Figure 3-d**), we observe a distinctly different response, as shown in **Figures 3-e** and **3-f**. The 8CB film forms defects; the FCDs, and the texture is similar to that achieved with deionized water and pure TB. This result indicates that the anchoring of the 8CB is planar when in contact with the TB solution containing *E. coli* cells.

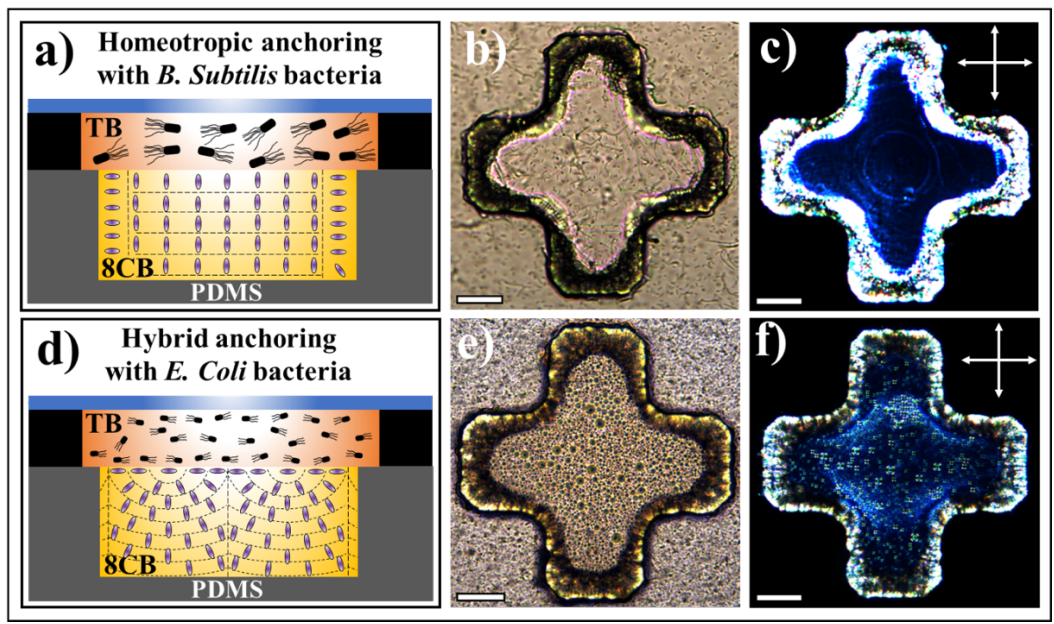

**Figure. 3** Interaction of the smectic interface with living microorganisms. (a) Side view sketch of the LC confined between PDMS and TB solution with *B. subtilis* cells. The Corresponding optical responses of the smectic LC under brightfield (b) and POM (c). These images indicate that the anchoring of the 8CB film is homeotropic. (d) Side view sketch of the LC confined between PDMS and TB with *E. coli* cells. The corresponding optical responses of the smectic LC under brightfield (e) and POM (f). These images indicate that the anchoring of the 8CB film is hybrid. The scale bars are 50 μm.

Based on the varying optical response of the LC film and the formation of defects, it is possible to conclude that the smectic interface may act as a biosensor to identify various types of bacteria. A LC biosensor operates on the principle that when specific biomolecules interact with the interface, they cause changes in the LC film's molecular arrangement and optical properties. These changes can be detected and analyzed, providing information about the presence, concentration, or activity of the target biological substances. Therefore, it is crucial to understand how the smectic responds differently to *B. subtilis* and *E. coli* bacteria in order to exploit this property for biosensing.

It is known that bacteria release a variety of proteins to invade and survive. Different types of bacteria secrete different proteins. For instance, previous studies have revealed that *B. subtilis* and *E. coli*, all strains combined, secrete at least 300[75] and 1600[76,77] proteins. This suggests that the difference in optical response with our system may be due to the variation in protein secretions. To prove this hypothesis, we conducted a proteomics study to determine what type of proteins *B. subtilis* and *E. coli* release during their growth (see **Tables S1 and S2** in **Supplementary Information)**. Our data show that the *B. subtilis* bacteria release approximately 77 proteins, whereas *E. coli* cells release 361 proteins. Interestingly, only 20 proteins were found to be common between the two microorganisms (see **Table S3** in **Supplementary Information**). In total, *B. subtilis* released 57 unique proteins, while *E. coli* released 341. These findings highlight the significant variation in protein composition between the two types of bacteria.

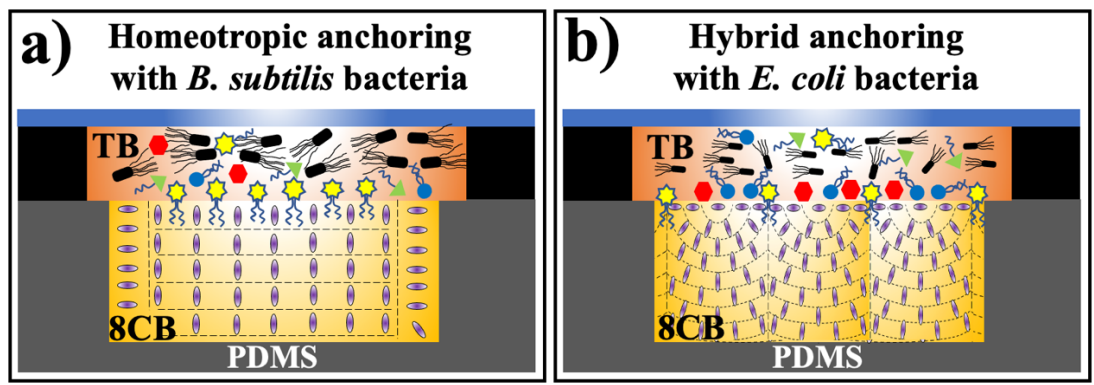

**Figure. 4** Adsorption of proteins at the smectic interface. (a) Sketch showing how proteins in a *B. subtilis* solution integrate their hydrophobic tails into the LC causing a homeotropic alignment of the 8CB molecules at the interface. (b) Sketch showing how proteins in an *E.coli* solution accumulate at the smectic interface causing the FCDs to decrease in size.

Previous groups have shown that *B. subtilis* bacteria release highly hydrophobic compounds,[78,79] while *E. coli* cells produce them in small quantities. These molecules are usually lipoproteins that tend to adsorb at the interface between the LC and the TB. Due to the hydrophobic nature of their hydrocarbon tails, the lipoproteins are integrated via their tails into the LC against the aqueous interface, causing a re-orientation of the LC molecules from planar to homeotropic.[47,80] This could explain the homeotropic alignment of the 8CB molecules in contact with the TB and *B. subtilis* bacteria, as shown in **Figure 4-a**. Our data show that *B. subtilis* produced only one type of lipoprotein, while *E. coli* produced seven (see **Tables S4 and S5** in **Supplementary Information**). Since only *B. subtilis* are inducing the homeotropic anchoring, we can then conclude that the *B. subtilis* cells tend to produce a more significant amount of hydrophobic compounds than *E. coli*, as confirmed in previous studies.[78,81] It is worth mentioning here that the proteomics technique helped us identify all the proteins secreted by both cells. However, it did not give us the precise amounts of these proteins. Although this study only provides qualitative information for differentiating between bacteria types, it can be reproduced.

Another significant result we noticed is that in the case of *E. coli* bacteria, the structure of FCDs at the smectic interface was different from one sample to another, particularly, their size changes (**Figure 2-e** and **Figure 3-e**). To better understand the origin of this variation, we investigated the sensitivity of our system to the concentration of *E. coli* cells ranging from zero to $3.056 \times 10^9$ cells/mL (prepared via dilutions). To a culture of *E. coli* grown overnight, we add sterile TB in the ratios of 1:9 (1 part of *E. coli* to 9 parts of sterile TB), 2:8, 4:6, 6:4, and 8:2 to dilute the bacterial concentration. The undiluted overnight culture was considered the highest concentration, and sterile TB was considered zero concentration. The densities of bacteria were measured by counting the number of bacteria in an image of 50 μm x 50 μm and extrapolating it to 1 mL volume for an approximate bacterial density per mL, as we know the thickness of the TB film, which is around 25 μm. The images were analyzed using FIJI to determine the average area of the defects. Here we neglect the defects that are less than 30 μm² because these defects are not stable and generally are not affected by the concentration of bacteria.

Optical images in **Figures 5-a to 5-f** show the formation of defects at all concentrations, but their size distribution varies as a function of *E. coli* density. These measurements indicate that the size of FCDs linearly decreases from ~ 85 μm² to 40 μm² with the concentration of *E. coli* changing from ~ 0 cells/mL to $3.1 \times 10^9$ cells/mL (**Figure 5-g**). This clearly indicates that either the bacteria secretions or their dynamics are contributing to the change in FCDs' size.

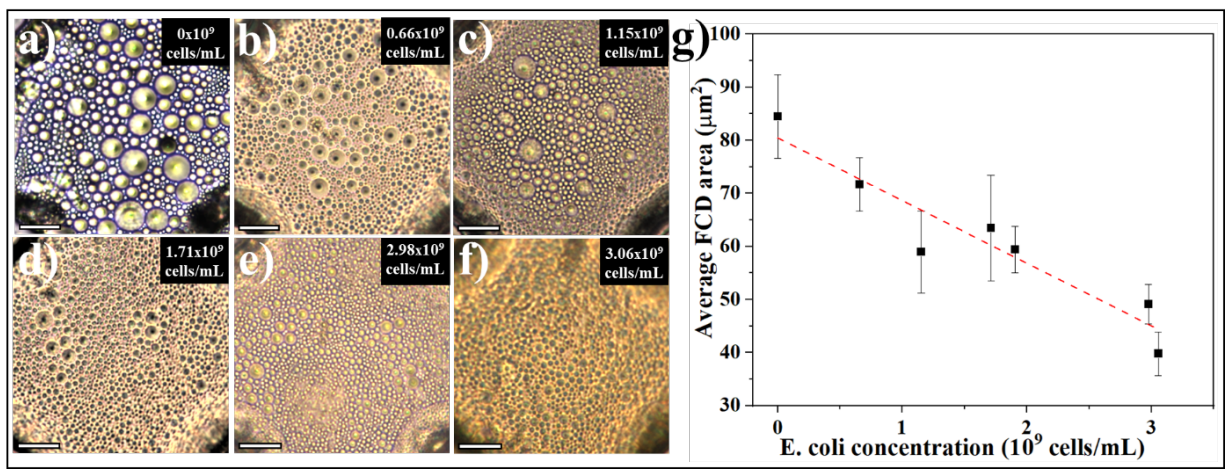

**Figure 5**. Effect of *E. coli* concentration on the size distribution of the FCD defects. (a-f) The size of the FCDs decreases with the concentration of *E. coli*. The scale bars are 25 μm. (g) Plot of the average FCD

area as a function of *E. coli* concentration. The bars represent the standard error. The dashed line is a linear fit of the data with the equation $y = -1 \times 10^{-8}x + 80.8$, and an $R^2 = 0.90$.

To clarify if the bacterial dynamics play a role in forming FCDs with different sizes, we repeated the same experiments with TB solutions containing only the secretions, prepared by removing the *E. coli* cells. A 10 mL overnight *E. coli* culture was pipetted into a 15 mL screw cap centrifuge tube and centrifuged at ~ 3250 g for 10 minutes. The supernatant containing bacterial secretory proteins was separated from the pellet and diluted by adding fresh and sterile TB. The dilutions prepared were 1/2, 1/4, 1/8, 1/16, and 1/32 of the supernatant concentration. The results obtained in **Figures 6-a to 6-f** were similar to those obtained with the solutions containing live *E. coli* (**Figure 5**). The FCDs' size decreases linearly with the concentration of the protein solution (**Figure 6-g**). Although the bacteria were absent, the TB solution containing only the proteins elicited a response similar to the TB solution containing the *E. coli* cell*s*. This indicates that bacterial dynamics have no role in creating defects. The physical presence of bacteria has negligible contribution, while bacterial secretions are responsible for the formation of FCDs and controlling their sizes. This also aligns with our hypothesis that a higher concentration of the proteins may stabilize the FCDs while restricting their expansion to a smaller size. The concentration of the secreted proteins is then forcing the FCDs to assume a smaller size by wrapping the smectic layers tighter to form a more significant number of smaller defects, as shown in **Figure 4-b**. We repeated the same experiment with *B. subtilis* secretions by removing the live cells. Similar results were obtained: The smectic anchoring switches from planar to homeotropic, confirming that the cells and their dynamics have a negligible effect on the properties of the smectic interface.

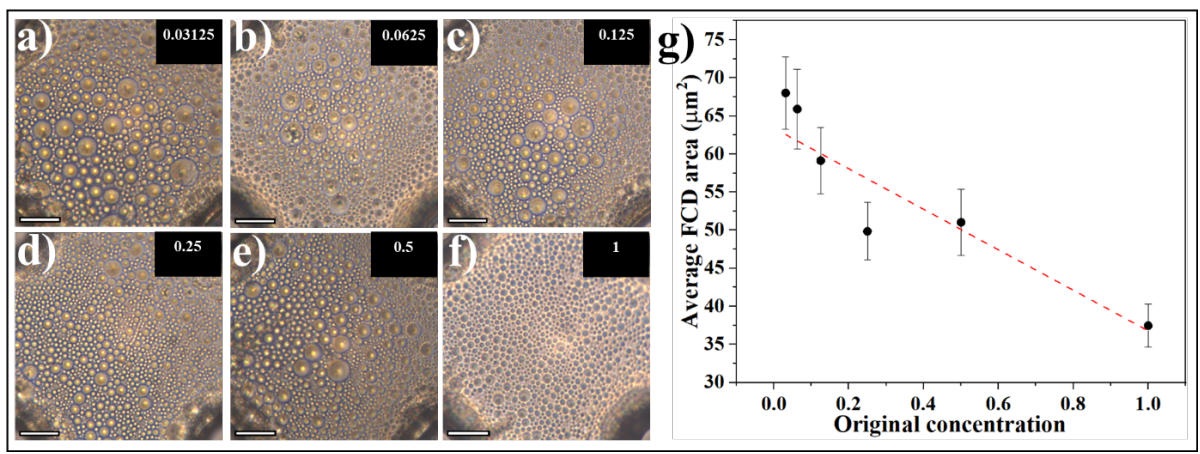

**Figure 6.** Effect of proteins released by *E. coli* cells on the size distribution of FCD defects. (a-f) The size of the FCDs decreases with the concentration of *E. coli* proteins. The scale bars are 25 μm. (g) Plot of the average FCD area as function of the *E-coli* protein concentration. The bars represent the standard error. The dashed line is a linear fit of the data with an equation $y = -28.60x + 64.61$, and an $R^2 = 0.86$.

All these results, demonstrate the potential of our smectic LC system to distinguish different types of living microorganisms due to the vast difference in their secreted proteins. They also show how smectic defects can be employed to estimate the concentration of the released biomolecules. These findings open doors for various applications in biomedical diagnostics, environmental monitoring, and the development of improved biosensors.

## 4. CONCLUSION

In this study, we successfully characterized the anchoring properties of smectic LCs in contact with *B. subtilis* and *E. coli* bacterial solutions using interfaces with known anchoring properties. Through our experiments, we demonstrated the ability to detect different types of bacteria by observing the changes in the optical properties of the smectic film. Furthermore, we elucidated the mechanism behind these changes by exploring the interactions between the proteins released by the bacteria and the smectic interfaces. Our findings also revealed the utility of topological defects at the smectic interface, specifically focal conic domains (FCDs), as a means to quantify the concentration of bacteria based on the amount of proteins present in the culture. These results highlight the promising potential of smectic LCs and their defects in biosensing applications. Overall, this work not only contributes to our understanding of the interactions between bacteria and smectic LCs but also paves the way for advancements in pathogen detection and protein-based sensing. Future research can focus on optimizing biosensor design, expanding the range of microorganisms examined, and exploring diverse applications in biomedical diagnostics, environmental monitoring, and other fields demanding rapid and sensitive biosensing capabilities.

## AUTHOR CONTRIBUTIONS

V.S.B and M.A.G. designed the experiments. V.S.B, M.A.G, and T.H.R.N performed the research, analyzed the data, and wrote the manuscript. All authors have read and agreed to the published version of the manuscript.

## CONFLICT OF INTEREST

The authors declare no conflict of interest.

**ACKNOWLEDGEMENTS**

This research was funded by the Joseph P. Healey Research Grant from the University of Massachusetts Boston. The authors would like to thank the Proteomics Core at the University of Massachusetts Boston for analyzing their samples on the Bruker tims-TOF HT.

**Supplementary Information**

**Topological defects at smectic interfaces as a potential tool for the biosensing of living microorganisms**

Vajra S. Badha,[a] Tagbo H.R. Niepa,[b,c] and Mohamed Amine Gharbi[a]*

[a] Department of Physics, University of Massachusetts Boston, Boston, MA 02125, USA

[b] Department of Chemical Engineering, Carnegie Mellon University, Pittsburgh, PA 15213, USA

[c] Department of Biomedical Engineering, Carnegie Mellon University, Pittsburgh, PA 15213, USA

* Corresponding author: Mohamed.Gharbi@umb.edu

# 1. Protocol of the Bicinchoninic Acid (BCA) Assay

To better understand the interaction between the smectic interface and *Escherichia coli* (*E. coli*) or *Bacillus subtilis* (*B. subtilis*) cells, and to investigate the underlying reason for the smectic LC's difference in response to different bacteria, we used a Bruker tims-TOF HT mass spectrometer to conduct a BCA proteomics assay and identify the total protein secretions in their solutions. After preparing the extracted protein samples, an aliquot of 100 μg of each sample was extracted and 20 μL of 500 mM ammonium bicarbonate was added to each sample, and the volumes were all brought up to 100 μL with Optima water, bringing the final concentration of ammonium bicarbonate to 100 mM. The samples were reduced with 2.1 μL of 500 mM dithiothreitol (DTT) for 45-mintes at 60 °C, and then cooled to room temperature before being alkylated with 11.5 μL of 500 mM iodoacetamide (IAA), in the dark, for 30-minutes. The samples were then digested with 2.5 μL of 1 mg/mL of Trypsin/Lys-C overnight at 37 °C.

*Liquid Chromatography with Ion Mobility – Mass Spectrometry (LC-IM-MS):* The samples were run on an Evosep One nLC (Evosep, Odense, Denmark) coupled to a Bruker timsTOF HT mass spectrometer (Bruker Scientific LLC, Billerica, MA) on a PepSep Endurance column (15 cm x 15 cm, 1.9 μm) (PepSep, Odense, Denmark) and operated in Parallel Accumulation – Serial Fragmentation (PASEF) mode with a scan range of 100-1700 m/z and a mobility range of 0.60-1.60 V·s/cm$^3$. The ramp time and accumulation times were both set to 100.0 ms, while the ramp rate to 9.42 Hz, and the MS averaging to 1. For the MS/MS parameters, the number of PASEF ramps was set to 10 with a total cycle time of 1.17 s with a target intensity of 10000 and the intensity threshold of 2500.

*Data Analysis:* Protein identification and quantification analysis were done with Parallel database Search Engine in Real-Time (PaSER, 2023, v 3.0, Bruker Scientific LLC, Billerica, MA) using ProLuCID,[1] DTASelect2[2,3] and Census.[4,5] Mass spectra were searched against Uniprot Escherichia_coli and Bacillus_subtilis up to date protein databases plus sequences of known contaminants such as keratin and porcine trypsin concatenated to a decoy database. TIMScore was appended to raw search results to use the peptide Collisional Cross Section (CCS) during the validation process.[6] These search results were validated, assembled, and filtered using the DTASelect program (version 2.1).

*Label Free Analysis:* A label-free quantitative analysis was performed using Census through PaSER (2023, v 3.0, Bruker Scientific LLC, Billerica, MA, http://www.bruker.com). The Census used protein identification results from DTASelect2 and generated a reconstructed MS1 based extracted ion chromatograms for each identified peptide. When peptides are not identified in all the relevant samples, Census went through spectra searching them using accurate precursor mass, retention time, ion mobility and charge states in order to retrieve them to build chromatograms.[7]

## 2. Results from the BCA assay

Below are the tables with the list of proteins identified in the BCA assay, from the *E. coli* and *B. subtilis* protein solutions obtained by removing the bacteria from the overnight cultures obtained.

**Table S1.** Total Proteins in the *E. coli* overnight growth solution

| | |
|---|---|
| 1 | Flagellin |
| 2 | Chaperone protein DnaK |
| 3 | 2,3,4,5-tetrahydropyridine-2,6-dicarboxylate N-succinyltransferase |
| 4 | 2,3-bisphosphoglycerate-dependent phosphoglycerate mutase (GN=gpmA) |
| 5 | 2,3-bisphosphoglycerate-independent phosphoglycerate mutase (GN=gpmI) |
| 6 | 2-dehydro-3-deoxyphosphooctonate aldolase |
| 7 | 2Fe-2S ferredoxin |
| 8 | 2-hydroxy-3-oxopropionate reductase |
| 9 | 30S ribosomal protein S1 OX=83333 |
| 10 | 30S ribosomal protein S1 OX=562 |
| 11 | 30S ribosomal protein S10 |
| 12 | 30S ribosomal protein S11 |
| 13 | 30S ribosomal protein S13 |
| 14 | 30S ribosomal protein S16 |
| 15 | 30S ribosomal protein S18 |
| 16 | 30S ribosomal protein S2 |
| 17 | 30S ribosomal protein S21 |
| 18 | 30S ribosomal protein S3 |
| 19 | 30S ribosomal protein S4 |
| 20 | 30S ribosomal protein S5 |
| 21 | 30S ribosomal protein S6 |
| 22 | 30S ribosomal protein S7 |
| 23 | 30S ribosomal protein S7 |
| 24 | 30S ribosomal protein S8 |
| 25 | 3-mercaptopyruvate sulfurtransferase |
| 26 | 3-oxoacyl-[acyl-carrier-protein] synthase 1 |
| 27 | 3-phenylpropionate dioxygenase beta subunit |
| 28 | 4-hydroxy-tetrahydrodipicolinate synthase |
| 29 | 50S ribosomal protein L1 |
| 30 | 50S ribosomal protein L10 |
| 31 | 50S ribosomal protein L14 GN=rplN |
| 32 | 50S ribosomal protein L14 GN=rplO |
| 33 | 50S ribosomal protein L15 |
| 34 | 50S ribosomal protein L16 |
| 35 | 50S ribosomal protein L17 |
| 36 | 50S ribosomal protein L19 |
| 37 | 50S ribosomal protein L2 |
| 38 | 50S ribosomal protein L21 |
| 39 | 50S ribosomal protein L24 |
| 40 | 50S ribosomal protein L25 |
| 41 | 50S ribosomal protein L27 |
| 42 | 50S ribosomal protein L28 |
| 43 | 50S ribosomal protein L3 |
| 44 | 50S ribosomal protein L31 |
| 45 | 50S ribosomal protein L32 |
| 46 | 50S ribosomal protein L4 |

| 47 | 50S ribosomal protein L5 |
| --- | --- |
| 48 | 50S ribosomal protein L6 |
| 49 | 50S ribosomal protein L7/L12 |
| 50 | 50S ribosomal protein L9 |
| 51 | 5-methyltetrahydropteroyltriglutamate--homocysteine methyltransferase |
| 52 | 5-methyltetrahydropteroyltriglutamate--homocysteine S-methyltransferase (Fragment) GN=D3C88_21865 |
| 53 | 5-methyltetrahydropteroyltriglutamate--homocysteine S-methyltransferase (Fragment) GN=E4K51_29170 |
| 54 | 6-phosphogluconate dehydrogenase, decarboxylating |
| 55 | Acetate OX=562 |
| 56 | Acetate kinase OX=83333 |
| 57 | Acetyl-coenzyme A carboxylase carboxyl transferase subunit alpha |
| 58 | Acid stress chaperone HdeA |
| 59 | Acid stress chaperone HdeB |
| 60 | Acidic protein MsyB |
| 61 | Aconitate hydratase B |
| 62 | Acyl carrier protein OX=562 |
| 63 | Acyl carrier protein O139:H28 |
| 64 | Adenylate kinase |
| 65 | Adenylosuccinate synthetase |
| 66 | ADP-L-glycero-D-manno-heptose-6-epimerase |
| 67 | Aerobic glycerol-3-phosphate dehydrogenase |
| 68 | Aldehyde-alcohol dehydrogenase |
| 69 | Alkyl hydroperoxide reductase C |
| 70 | Alkyl hydroperoxide reductase subunit F |
| 71 | Aminoimidazole riboside kinase |
| 72 | Aminotransferase class III-fold pyridoxal phosphate-dependent enzyme |
| 73 | Anaerobic glycerol-3-phosphate dehydrogenase subunit A |
| 74 | Anaerobic glycerol-3-phosphate dehydrogenase subunit B |
| 75 | Anaerobic glycerol-3-phosphate dehydrogenase subunit B |

| 76 | Anaerobic glycerol-3-phosphate dehydrogenase subunit C |
| --- | --- |
| 77 | Anti-sigma-28 factor FlgM (Fragment) |
| 78 | Asparagine--tRNA ligase |
| 79 | Aspartate aminotransferase |
| 80 | Aspartate ammonia-lyase |
| 81 | Aspartate-semialdehyde dehydrogenase |
| 82 | ATP synthase subunit alpha |
| 83 | ATP-cone domain-containing protein |
| 84 | ATP-dependent Clp protease ATP-binding subunit ClpX |
| 85 | ATP-dependent protease ATPase subunit HslU |
| 86 | ATP-dependent protease subunit HslV |
| 87 | Autonomous glycyl radical cofactor |
| 88 | Bacterial non-heme ferritin |
| 89 | Bacterioferritin |
| 90 | Basal-body rod modification protein FlgD OX=562 |
| 91 | Basal-body rod modification protein FlgD OX=83333 |
| 92 | Bifunctional aldehyde-alcohol dehydrogenase AdhE |
| 93 | Bifunctional aspartokinase/homoserine dehydrogenase 1 |
| 94 | Bifunctional NADP-dependent 3-hydroxy acid dehydrogenase/3-hydroxypropionate dehydrogenase YdfG |
| 95 | Catalase-peroxidase |
| 96 | Cell division protein FtsZ |
| 97 | Cell division protein ZapB |
| 98 | Cell shape-determining protein MreB |
| 99 | Chaperedoxin |
| 100 | Chaperone protein ClpB |
| 101 | Chaperone protein DnaK (Heat shock protein 70) (Heat shock 70 kDaprotein) (HSP70) |
| 102 | Chaperone protein DnaK (Heat shock protein 70) (Heat shock 70 kDaprotein) (HSP70) |
| 103 | Chaperone protein HtpG |
| 104 | Chaperone protein Skp |
| 105 | Chaperonin GroEL |
| 106 | Chaperonin GroEL 1 |
| 107 | Chemotaxis protein CheW |
| 108 | Chemotaxis protein CheY |
| 109 | Citrate synthase |
| 110 | Co-chaperonin GroES |
| 111 | Cold shock-like protein CspC |
| 112 | Cold shock-like protein CspE |
| 113 | CTP synthase |
| 114 | Curved DNA-binding protein |
| 115 | Cysteine desulfurase IscS |
| 116 | Cysteine synthase |
| 117 | Cystine transporter subunit |
| 118 | Cytidine deaminase |
| 119 | Cytochrome bd-I ubiquinol oxidase subunit 1 |
| 120 | Cytosol non-specific dipeptidase |
| 121 | Deoxyribose-phosphate aldolase OX=316385 |
| 122 | deoxyribose-phosphate aldolase OX=562 |
| 123 | D-galactose/methyl-galactoside binding periplasmic protein MglB |
| 124 | Dihydrolipoamide acetyltransferase component of pyruvate dehydrogenase complex |
| 125 | Dihydrolipoyl dehydrogenase |
| 126 | Dihydrolipoyllysine-residue acetyltransferase component of pyruvate dehydrogenase complex |
| 127 | Dihydrolipoyllysine-residue succinyltransferase component of 2-oxoglutarate dehydrogenase complex |
| 128 | Dihydroxyacetone kinase subunit K |
| 129 | Dimethyl sulfoxide reductase DmsA |
| 130 | Dipeptide-binding protein |
| 131 | DNA gyrase subunit B |
| 132 | DNA protection during starvation protein |
| 133 | DNA-binding protein HU-alpha |
| 134 | DNA-binding protein HU-beta |
| 135 | DNA-binding transcriptional dual regulator CRP |
| 136 | DNA-directed RNA polymerase subunit alpha |
| 137 | DNA-directed RNA polymerase subunit beta O1:K1 |
| 138 | DNA-directed RNA polymerase subunit beta OX=562 |
| 139 | DNA-directed RNA polymerase subunit beta' O1:K1 |
| 140 | DNA-directed RNA polymerase subunit beta' OX=562 |
| 141 | DNA-directed RNA polymerase subunit omega |
| 142 | D-tagatose-1,6-bisphosphate aldolase subunit GatY |
| 143 | D-tagatose-1,6-bisphosphate aldolase subunit GatZ |
| 144 | DUF2786 domain-containing protein |
| 145 | Ecotin |
| 146 | Elongation factor 4 |
| 147 | Elongation factor G |
| 148 | Elongation factor Ts |
| 149 | Elongation factor Tu |
| 150 | Elongation factor Tu (Fragment) |
| 151 | Elongation factor Tu 1 |
| 152 | Elongation protein Tu GTP binding domain-containing protein (Fragment) |
| 153 | Enolase |
| 154 | Enoyl-[acyl-carrier-protein] reductase [NADH] FabI |
| 155 | Entericidin B |
| 156 | Exonuclease SbcC |
| 157 | Fe/S biogenesis protein NfuA |
| 158 | Fimbrial family protein |
| 159 | FKBP-type 22 kDa peptidyl-prolyl cis-trans isomerase |
| 160 | FKBP-type peptidyl-prolyl cis-trans isomerase FkpA |
| 161 | FKBP-type peptidyl-prolyl cis-trans isomerase SlyD |
| 162 | Flagellar basal body rod protein FlgB |
| 163 | Flagellar hook protein FlgE |
| 164 | Flagellar hook-associated protein 1 OX=562 |
| 165 | Flagellar hook-associated protein 1 OX=83333 |
| 166 | Flagellar hook-associated protein 1 OX=562 |
| 167 | Flagellar hook-associated protein 2 |
| 168 | Flagellar hook-associated protein 3 OX=83333 |
| 169 | Flagellar hook-associated protein 3 OX=562 |

| 170 | Flagellin |
| --- | --- |
| 171 | Flagellin (Fragment) GN=fliC |
| 172 | Flagellin (Fragment) GN=fllA55 |
| 173 | Flagellin (Fragment) GN=flnA |
| 174 | Flagellin (Fragment) GN=fliC |
| 175 | Flagellin FliC |
| 176 | Formate acetyltransferase 1 |
| 177 | Fructose-1,6-bisphosphatase class 1 |
| 178 | Fructose-bisphosphate aldolase class 1 |
| 179 | Fructose-bisphosphate aldolase class 2 |
| 180 | Fumarate hydratase class I |
| 181 | Fumarate hydratase class I, anaerobic |
| 182 | Fumarate reductase flavoprotein subunit OX=83333 |
| 183 | Fumarate reductase flavoprotein subunit OX=562 |
| 184 | GapA (Fragment) |
| 185 | Glucose-1-phosphatase |
| 186 | Glucose-6-phosphate isomerase |
| 187 | Glutamate decarboxylase D3G36_07530 |
| 188 | glutamate decarboxylase GN=CCV24_003315 |
| 189 | Glutamate decarboxylase (Fragment) GN=ACN68_06885 |
| 190 | glutamate decarboxylase (Fragment) GN=DTM45_28360 |
| 191 | glutamate decarboxylase (Fragment) GN=GNW61_08045 |
| 192 | glutamate decarboxylase (Fragment) GN=FPI65_32745 |
| 193 | Glutamate decarboxylase (Fragment) GN=ELX76_24345 |
| 194 | Glutamate decarboxylase alpha |
| 195 | Glutamate/gamma-aminobutyrate antiporter |
| 196 | Glutamate--tRNA ligase |
| 197 | Glutamine-binding periplasmic protein |
| 198 | Glutaredoxin 2 |
| 199 | Glutaredoxin 3 |
| 200 | Glutaredoxin 4 |
| 201 | Glyceraldehyde-3-phosphate dehydrogenase A |
| 202 | Glycerol kinase |
| 203 | Glycerol-3-phosphate transporter |
| 204 | Glycerophosphodiester phosphodiesterase |
| 205 | Glycerophosphodiester phosphodiesterase, periplasmic |
| 206 | Glycine--tRNA ligase beta subunit |
| 207 | Glycogen synthase |
| 208 | Glyoxalase ElbB |
| 209 | HAMP domain-containing protein (Fragment) |
| 210 | High-affinity zinc uptake system protein ZnuA |
| 211 | Histidine ABC transporter substrate-binding protein HisJ |
| 212 | Histidine phosphatase family protein |
| 213 | HlyD-family secretion protein |
| 214 | Hydrogenase-1 large chain |
| 215 | Inorganic pyrophosphatase |
| 216 | Inosine-5'-monophosphate dehydrogenase |
| 217 | Iron-containing alcohol dehydrogenase |
| 218 | Isocitrate dehydrogenase [NADP] OX=83333 |
| 219 | Isocitrate dehydrogenase [NADP] (Fragment) OX=562 |
| 220 | Isocitrate dehydrogenase [NADP] (Fragment) OX=562 |
| 221 | Isocitrate lyase |
| 222 | Isoleucine--tRNA ligase |
| 223 | KHG/KDPG aldolase |
| 224 | L-asparaginase 2 |
| 225 | L-cystine-binding protein TcyJ |
| 226 | Leu/Ile/Val-binding protein |
| 227 | Leucine--tRNA ligase |
| 228 | L-fucose mutarotase |
| 229 | Lipid A core - O-antigen ligase and related enzymes |
| 230 | Lon protease |
| 231 | Lysine/arginine/ornithine ABC transporter substrate-binding protein ArgT |
| 232 | Lysine--tRNA ligase OX=83333 |
| 233 | Lysine--tRNA ligase O6:K15:H31 |
| 234 | Lysine--tRNA ligase, heat inducible |

| 235 | Lysozyme |
| --- | --- |
| 236 | Major outer membrane lipoprotein Lpp |
| 237 | Malate dehydrogenase |
| 238 | Maltodextrin-binding protein (Fragment) |
| 239 | Maltose/maltodextrin-binding periplasmic protein |
| 240 | Mannitol-1-phosphate 5-dehydrogenase |
| 241 | Metal-binding protein ZinT |
| 242 | Methyl-accepting chemotaxis protein II |
| 243 | Methylated-DNA--[protein]-cysteine S-methyltransferase |
| 244 | Molecular chaperone |
| 245 | Molybdate-binding periplasmic protein |
| 246 | Molybdate-binding protein ModA |
| 247 | Molybdenum cofactor biosynthesis protein B |
| 248 | Molybdopterin guanine dinucleotide synthesis B family protein |
| 249 | Multidrug efflux pump subunit AcrA |
| 250 | NAD(P)H dehydrogenase (quinone) |
| 251 | Negative regulator of flagellin synthesis OX=562 |
| 252 | Negative regulator of flagellin synthesis OX=83333 |
| 253 | NH(3)-dependent NAD(+) synthetase |
| 254 | Oligopeptide ABC transporter substrate-binding protein |
| 255 | Outer membrane lipoprotein DolP |
| 256 | Outer membrane lipoprotein Slp |
| 257 | Outer membrane protein A |
| 258 | Outer membrane protein Slp |
| 259 | Outer-membrane lipoprotein carrier protein |
| 260 | Oxygen-insensitive NAD(P)H nitroreductase |
| 261 | PEP-dependent dihydroxyacetone kinase, ADP-binding subunit DhaL |
| 262 | PEP-dependent dihydroxyacetone kinase, dihydroxyacetone-binding subunit DhaK |
| 263 | PEP-dependent dihydroxyacetone kinase, phosphoryl donor subunit DhaM |
| 264 | Peptidoglycan-associated lipoprotein |
| 265 | PerC family transcriptional regulator |

| 266 | Peroxiredoxin |
| --- | --- |
| 267 | Phosphate acetyltransferase |
| 268 | Phosphocarrier protein HPr |
| 269 | Phosphoenolpyruvate carboxykinase (ATP) O1:K1 |
| 270 | Phosphoenolpyruvate carboxykinase (ATP) OX=562 |
| 271 | Phosphoenolpyruvate-protein phosphotransferase |
| 272 | Phosphoglycerate kinase |
| 273 | Phosphoglycerate mutase (2,3-diphosphoglycerate-independent) (Fragment) |
| 274 | Phosphopentomutase |
| 275 | Polyribonucleotide nucleotidyltransferase |
| 276 | Potassium binding protein Kbp |
| 277 | Protein disaggregation chaperone |
| 278 | Protein FlxA |
| 279 | Protein GrpE |
| 280 | Protein transport protein HofC |
| 281 | Protein YciN |
| 282 | Protein YdgH |
| 283 | Protein YgiW |
| 284 | protein-secreting ATPase (Fragment) |
| 285 | PTS system glucose-specific EIIA component |
| 286 | PTS system mannose-specific EIIAB component |
| 287 | Purine nucleoside phosphorylase DeoD-type |
| 288 | Putative Fe-S oxidoreductases (SAM domain protein) |
| 289 | Putative glucose-6-phosphate 1-epimerase |
| 290 | Putative monooxygenase YdhR |
| 291 | Putative NAD(P)H nitroreductase YdjA |
| 292 | Putative selenoprotein YdfZ |
| 293 | Pyruvate dehydrogenase E1 component |
| 294 | Pyruvate kinase I |
| 295 | Pyruvate kinase II |
| 296 | Respiratory nitrate reductase 1 alpha chain |
| 297 | Ribonuclease E |
| 298 | Ribose import binding protein RbsB |

| 299 | ribose-5-phosphate isomerase (Fragment) |
| --- | --- |
| 300 | Ribose-phosphate pyrophosphokinase |
| 301 | Ribosome-associated inhibitor A |
| 302 | Ribosome-recycling factor |
| 303 | RNA polymerase-binding transcription factor DksA |
| 304 | Sec translocon accessory complex subunit YajC |
| 305 | Selenide, water dikinase |
| 306 | Septum site-determining protein MinD |
| 307 | Serine--tRNA ligase |
| 308 | Small-conductance mechanosensitive channel |
| 309 | Stringent starvation protein A |
| 310 | Substrate-binding domain-containing protein (Fragment) |
| 311 | succinate dehydrogenase |
| 312 | Superoxide dismutase [Fe] |
| 313 | Thiol peroxidase |
| 314 | Thioredoxin 1 |
| 315 | Thiosulfate-binding protein |
| 316 | Threonine--tRNA ligase |
| 317 | TIGR03756 family integrating conjugative element protein |
| 318 | Transaldolase GN=tal2 |
| 319 | Transaldolase GN=tal |
| 320 | Transaldolase OX=679206 |
| 321 | Transcription elongation factor GreA |
| 322 | Transcription termination/antitermination protein NusA |
| 323 | Transcription termination/antitermination protein NusG |
| 324 | Transketolase GN=tktA_3 |
| 325 | Transketolase GN=tktA_1 |
| 326 | Transketolase (Fragment) |
| 327 | Transketolase 1 |
| 328 | Translation elongation factor G |
| 329 | Translation initiation factor IF-1 |
| 330 | Translation initiation factor IF-3 |
| 331 | Trehalose-6-phosphate hydrolase |
| 332 | Trigger factor O1:K1 |
| 333 | Trigger factor OX=562 |
| 334 | Triosephosphate isomerase |
| 335 | tRNA-dihydrouridine synthase B |
| 336 | Tryptophanase O139:H28 |
| 337 | Tryptophanase OX=344610 |
| 338 | Uncharacterized lipoprotein YbaY |
| 339 | Uncharacterized protein GN=C9E67_19705 |
| 340 | Uncharacterized protein GN=EL79_5186 |
| 341 | Uncharacterized protein GN=BANRA_05067 |
| 342 | Uncharacterized protein OX=1268991 GN=HMPREF1604_00196 |
| 343 | Uncharacterized protein (Fragment) |
| 344 | Uncharacterized protein Yah |
| 345 | Uncharacterized protein YccJ |
| 346 | Uncharacterized protein YjeI |
| 347 | Uncharacterized protein YncE |
| 348 | Uncharacterized protein YnfD |
| 349 | Uncharacterized protein YqjD |
| 350 | Universal stress protein F |
| 351 | UPF0149 protein YgfB |
| 352 | UPF0227 protein YcfP |
| 353 | UPF0234 protein YajQ |
| 354 | UPF0304 protein YfbU |
| 355 | UPF0325 protein YaeH |
| 356 | UPF0381 protein YfcZ |
| 357 | UPF0434 protein YcaR |
| 358 | Uracil phosphoribosyltransferase |
| 359 | Uridine phosphorylase OX=83333 |
| 360 | Uridine phosphorylase OX=562 |
| 361 | YgiW/YdeI family stress tolerance OB fold protein |

**Table S2.** Total Proteins in the *B. subtilis* overnight growth solution

| 1 | (R,R)-butanediol dehydrogenase |
|---|---|
| 2 | 2-hydroxy-3-keto-5-methylthiopentenyl-1-phosphate phosphatase |
| 3 | 30S ribosomal protein S1 homolog |
| 4 | 30S ribosomal protein S2 |
| 5 | 30S ribosomal protein S7 |
| 6 | Acetolactate synthase |
| 7 | Adenylate kinase |
| 8 | Alkyl hydroperoxide reductase C |
| 9 | Asparagine synthetase [glutamine-hydrolyzing] 1 |
| 10 | ATP synthase subunit alpha |
| 11 | Cell wall-associated protease |
| 12 | Chaperone protein DnaK |
| 13 | Cold shock protein CspD |
| 14 | Cryptic catabolic NAD-specific glutamate dehydrogenase GudB |
| 15 | D-alanyl carrier protein |
| 16 | DNA gyrase subunit A |
| 17 | DNA processing protein DprA |
| 18 | DNA-binding protein HU 1 |
| 19 | Ferredoxin |
| 20 | Flagellar M-ring protein |
| 21 | Flagellin |
| 22 | Glutamate synthase [NADPH] large chain |
| 23 | Glutamine synthetase |
| 24 | Glycerol kinase |
| 25 | Heme-degrading monooxygenase HmoB |
| 26 | Immunity protein YezG |
| 27 | Inosine-5'-monophosphate dehydrogenase |
| 28 | Isoleucine--tRNA ligase |
| 29 | Lactate utilization protein C |
| 30 | L-Ala-D/L-Glu epimerase |
| 31 | Malate dehydrogenase |
| 32 | Methionine-binding lipoprotein MetQ |
| 33 | NAD-dependent malic enzyme 1 |
| 34 | Negative regulator of genetic competence ClpC/MecB |
| 35 | Nuclease SbcCD subunit C |
| 36 | Ornithine aminotransferase |
| 37 | Penicillin-binding protein 1A/1B |
| 38 | Phage-like element PBSX protein XkdF |
| 39 | Phosphoglycerate kinase |
| 40 | Polyketide synthase PksJ |
| 41 | Polyribonucleotide nucleotidyltransferase |
| 42 | Probable cytosol aminopeptidase |
| 43 | Protein translocase subunit SecA |
| 44 | Purine nucleoside phosphorylase 1 |
| 45 | Putative cell wall shaping protein YabE |
| 46 | Putative cytochrome P450 YjiB |
| 47 | Putative nitrogen fixation protein YutI |
| 48 | Putative tRNA-binding protein YtpR |
| 49 | Pyruvate dehydrogenase E1 component subunit alpha |
| 50 | Ribonucleoside-diphosphate reductase subunit beta |
| 51 | Ribosome biogenesis GTPase A |
| 52 | Ribosome-recycling factor |
| 53 | S-adenosylmethionine synthase |
| 54 | Serine protease Do-like HtrA |
| 55 | SPbeta prophage-derived stress response protein SCP1 |
| 56 | Sporulation kinase A |
| 57 | Succinate dehydrogenase flavoprotein subunit |
| 58 | Superoxide dismutase [Mn] |
| 59 | Thioredoxin |
| 60 | Toxin YqcG |
| 61 | Transcription termination/antitermination protein NusA |
| 62 | Transcription termination/antitermination protein NusG |
| 63 | Transcriptional regulatory protein ResD |

| 64 | Trifunctional nucleotide phosphoesterase protein YfkN |
| 65 | Trigger factor |
| 66 | tRNA nuclease WapA |
| 67 | tRNA-2-methylthio-N(6)-dimethylallyladenosine synthase |
| 68 | Uncharacterized phosphotransferase YvkC |
| 69 | Uncharacterized protein YceE |
| 70 | Uncharacterized protein YncM |
| 71 | Uncharacterized protein YneR |
| 72 | Uncharacterized protein YppF |
| 73 | Uncharacterized protein YqjE |
| 74 | Uncharacterized protein YwoF |
| 75 | UPF0702 transmembrane protein YdfS |
| 76 | Vegetative catalase |
| 77 | Vegetative protein 296 |

**Table S3.** Proteins identified in both the *E. coli* and *B. subtilis* overnight growth solution

| 1 | 30S ribosomal protein S1 homolog |
| 2 | 30S ribosomal protein S7 |
| 3 | Adenylate kinase |
| 4 | Alkyl hydroperoxide reductase C |
| 5 | ATP synthase subunit alpha |
| 6 | Chaperone protein DnaK |
| 7 | Cold shock protein CspD |
| 8 | Flagellin |
| 9 | Glycerol kinase |
| 10 | Inosine-5'-monophosphate dehydrogenase |
| 11 | Isoleucine--tRNA ligase |
| 12 | Malate dehydrogenase |
| 13 | Phosphoglycerate kinase |
| 14 | Polyribonucleotide nucleotidyltransferase |
| 15 | Pyruvate dehydrogenase E1 component subunit alpha |
| 16 | Ribosome-recycling factor |

| 17 | Thioredoxin |
| --- | --- |
| 18 | Transcription termination/antitermination protein NusA |
| 19 | Transcription termination/antitermination protein NusG |
| 20 | Trigger factor |

**Table S4.** Lipoproteins identified in *E. coli* solution.

| 1 | Major outer membrane lipoprotein Lpp |
| --- | --- |
| 2 | Outer membrane lipoprotein DolP |
| 3 | Outer membrane lipoprotein Slp 1 |
| 4 | Outer membrane lipoprotein Slp 2 |
| 5 | Outer-membrane lipoprotein carrier protein |
| 6 | Peptidoglycan-associated lipoprotein |
| 7 | Uncharacterized lipoprotein YbaY |

**Table S5.** Lipoproteins identified in *B. subtilis* solution.

| 1 | Methionine-binding lipoprotein MetQ |
| --- | --- |